\documentclass[letterpaper,twocolumn,10pt]{article}
\usepackage{usenix2019_v3}

\usepackage{tikz}
\usepackage{amsmath}

\usepackage{filecontents}

\newcommand{\why}{\textbf{\textit{why}}}
\newcommand{\where}{\textbf{\textit{where}}}
\newcommand{\how}{\textbf{\textit{how}}}

\begin{document}

\date{}

\title{\Large \bf Privacy Aspects of Provenance Queries}

\author{
{\rm Tanja Auge}\\
University of Rostock
\and
{\rm Nic Scharlau}\\
University of Rostock
\and
{\rm Andreas Heuer}\\
University of Rostock
} 

\maketitle

\begin{abstract}
Given a query result of a big database, \why-provenance can be used to calculate the necessary part of this database, consisting of so-called witnesses. If this database consists of personal data, privacy protection has to prevent the publication of these witnesses.
This implies a natural conflict of interest between publishing original data (\textit{provenance}) and protecting these data (\textit{privacy}).

In this paper, privacy goes beyond the concept of personal data protection. The paper gives an extended definition of privacy as intellectual property protection. If the provenance information is not sufficient to reconstruct a query result, additional data such as witnesses or provenance polynomials have to be published to guarantee traceability. 
Nevertheless, publishing this provenance information might be a problem if (significantly) more tuples than necessary can be derived from the original database.
At this point, it is already possible to violate privacy policies, provided that quasi identifiers are included in this provenance information. With this poster, we point out fundamental problems and discuss first proposals for solutions. 
\end{abstract}

\section{Privacy vs. Provenance}
For us, the term privacy goes beyond the concept of (mostly personal) data protection. Rather, we mean the protection of data in general. Reasons for the protection of data can be economic ones (intellectual properties), since generating such data is often very time-consuming and expensive. The identification of personal or internal company information should also be strictly prevented. 

\begin{figure}
\centering
\includegraphics[scale=0.23]{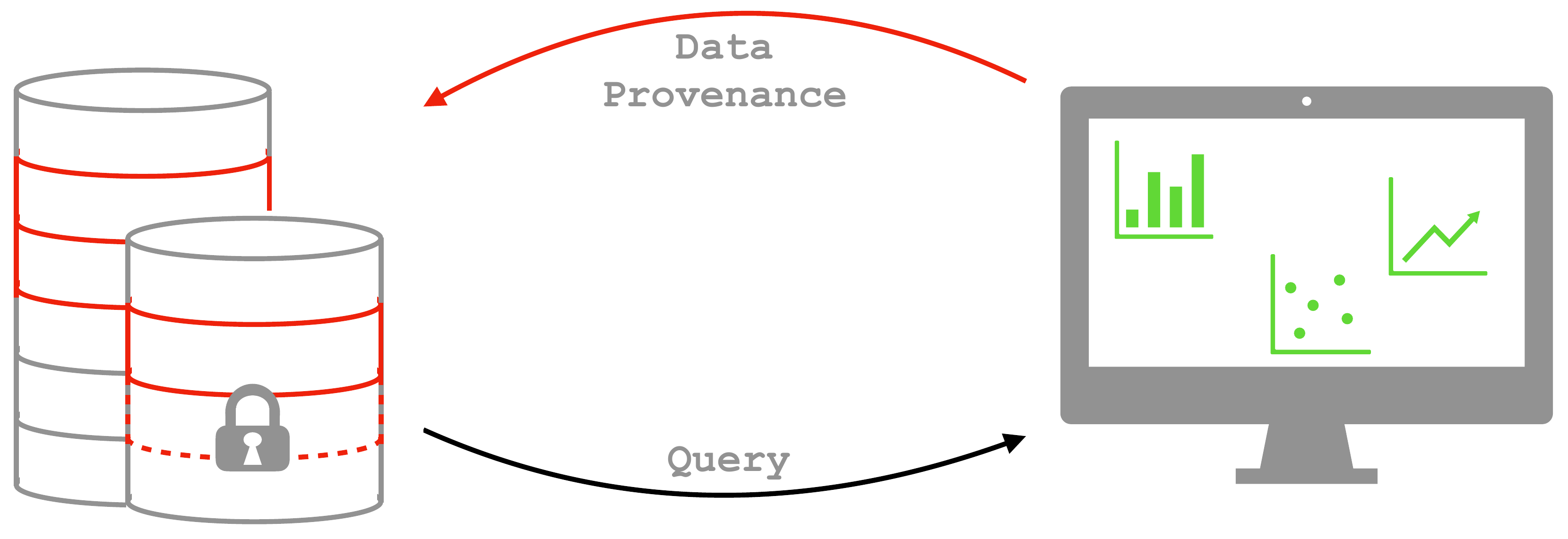}
\caption{Motivation}
\label{fig:motivation}
\end{figure}

Since queries that occur in the context of projects can become arbitrarily complex -- simple selections and projections, even joins and aggregations -- the inversion of these queries is often not 100$\%$ possible or necessary. However, using provenance enables us to perform this inversion as accurately as possible \cite{AH19}. This implies a natural conflict of interest between publishing original data (\textit{provenance}) and protecting these data (\textit{privacy}), and it may be possible to reconstruct parts of the source instance that contradict the privacy concepts (see Figure \ref{fig:motivation}). 

\textbf{Privacy} refers to the protection of (personal) data against unauthorized collection, storage, and publication. 
This becomes difficult if even the combination of apparently harmless attributes (\textit{quasi-identifiers}) can lead to a clear identification.

\textbf{Data Provenance} is concerned with the origin of a data analysis. This analysis may be a database query using aggregation or selection. We distinguish three provenance-questions answered by using the names of the source relations (\where), witness bases (\why, \cite{BKT01}), or provenance polynomials (\how, \cite{GT17}), as well as additional information which are sometimes necessary for reconstructing concrete lost attribute values. 

\textbf{Privacy and Provenance} conflicts have already been discussed in various papers.
For example, a formalization of security properties such as disclosure and obfuscation is shown in \cite{Che11}.
Besides Data Provenance, there also exist other types of provenance we need to face. At the workflow level, privacy and provenance have already been investigated by \cite{DKRSTC11}. 

\begin{figure*}[t]
\centering
\includegraphics[width=0.8275\textwidth]{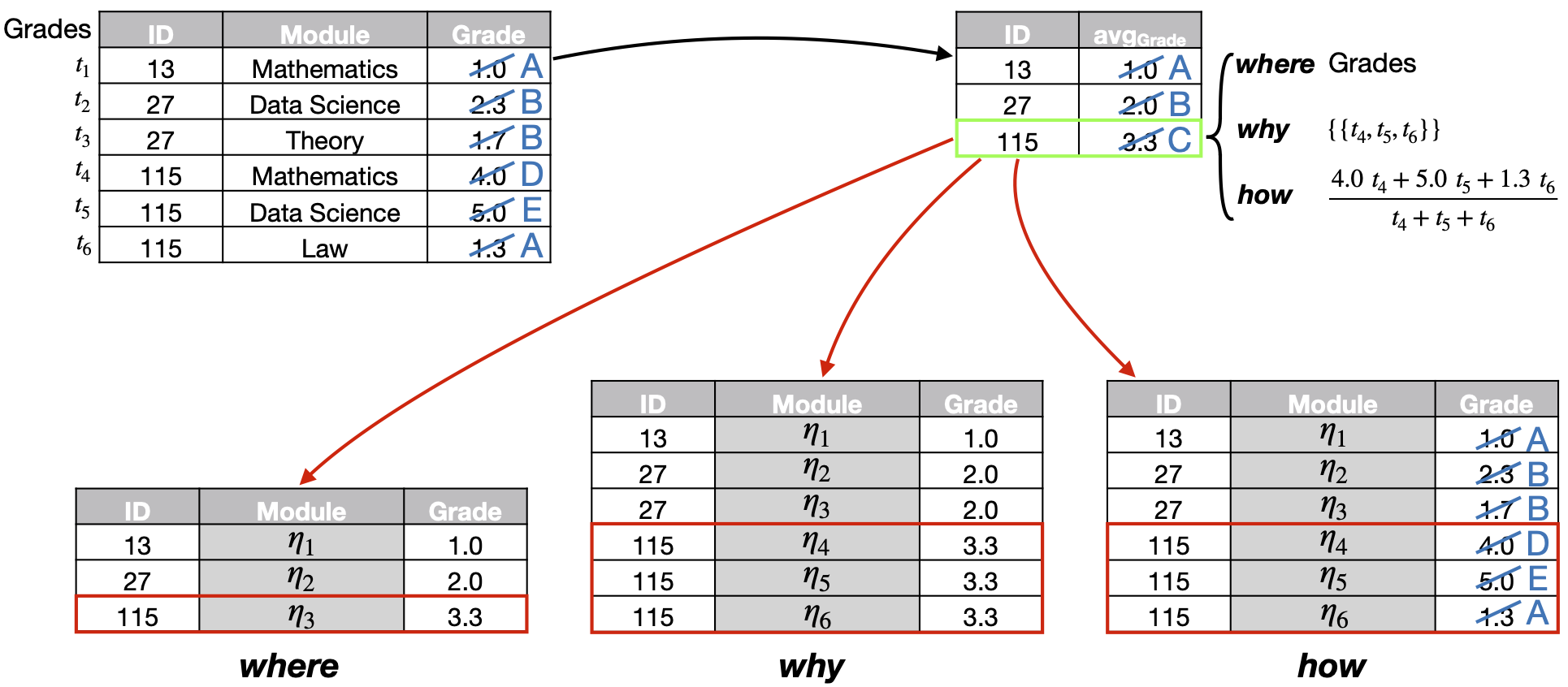}
\caption{Possible provenance-based database reconstructions (red) including generalization (blue) as solution approach}
\label{pic:PrivacyProvenanceMainIdea}
\end{figure*}

\section{Problems using \where, \why ~and \how}
Data provenance can have different characteristics, generating different provenance information and therefore different privacy problems. 

Let us imagine a data set with personal data such as a university database. It contains data about its students and staff, equipment and buildings, and much more. A sensitive attribute w.r.t.\ privacy aspects is the \texttt{Grade} attribute in the \texttt{Grades} relation storing the results of the exams. Instead of extracting the exact grade for every exam, a query calculating the average grade per student should be allowed to be performed by the university administration. Different sub-databases can now be calculated depending on the choice of stored provenance. Figure \ref{pic:PrivacyProvenanceMainIdea} shows the result tables extended to the original schema using \where-provenance (left), the extension to the exact number of tuples when using \why-provenance (center), or the representation of all individual grades per student in the case of \how-provenance (right). In all situations the \texttt{Module} attribute can not be reconstructed. Hence, modules are reconstructed introducing null values (highlighted in gray).

These data can be reconstructed by \where-, \why- and \how-provenance with the techniques described in \cite{AH19}. We will now consider the privacy aspects of these results of the provenance queries: (1) Using \where, there is generally not enough data worth protecting and reproducibility of the data is not guaranteed. Data protection aspects are therefore negligible. If we interpret \textbf{\textit{where}} as tuple names and we save not only the scheme but the tuple itself, this can lead to major privacy problems. However, this second \where \ approach is subject of our current work. (2) In the case of \why-provenance, we may encounter privacy problems, if the variance of the distribution of attribute values is equal to zero. However, this only applies for special cases not known to the user interpreting the results of the provenance queries. (3) \textbf{\textit{How}} often calculates too much recoverable information, so that privacy aspects are likely to be a major problem with this technology. 
\section{Possible Solutions to the Privacy Problem}
For solving the problems generated by the different provenance queries, we examined different approaches such as generalization and suppression, differential privacy, permutation of attribute values, and intensional (instead of extensional) answers to provenance queries \cite{Sch20}. 

Intensional provenance answers represent one solution approach to combine data provenance and privacy. It can be realized, e.g., by the generalization of attribute values. The idea of generalization is shown in Figure \ref{pic:PrivacyProvenanceMainIdea} by generalizing the grade from a concrete number like $1.0$ or $1.3$ to a grade area of $A$ (highlighted in blue). This results in a (hopefully acceptable) loss of information, while approaching a solution to the privacy problem of protecting sensitive attribute values.

\section*{Acknowledgments}
We thank Goetz Graefe for his support and comments during the development of this work.

\small
\bibliographystyle{plain}
\bibliography{\jobname}

\end{document}